%Paper: astro-ph/9410036
%From: egblackm@cfa160.harvard.edu (Eric Blackman)
%Date: Wed, 12 Oct 94 13:25:16 EDT

\def\.{\mathaccent 95}

\def\ga{\gamma}

\def\ka{\kappa}
\def\la{\lambda}

\def\si{\sigma}

\def\Ga{\Gamma}

\def\frac#1#2{{\textstyle{{#1}\over {#2}}}}

\def\lsim{\mathrel{\rlap{\lower4pt\hbox{\hskip1pt$\sim$}}
    \raise1pt\hbox{$<$}}}
\def\gsim{\mathrel{\rlap{\lower4pt\hbox{\hskip1pt$\sim$}}
    \raise1pt\hbox{$>$}}}
\def\sqr#1#2{{\vcenter{\vbox{\hrule height.#2pt
         \hbox{\vrule width.#2pt height#1pt \kern#1pt
         \vrule width.#2pt}
         \hrule height.#2pt}}}}

% Next 5 lines define \lapprox and \gapprox: "less than or approximately
% equal to" and "greater than or approximately equal to".
\newbox\grsign \setbox\grsign=\hbox{$>$} \newdimen\grdimen \grdimen=\ht\grsign
\newbox\simlessbox \newbox\simgreatbox
\setbox\simgreatbox=\hbox{\raise.5ex\hbox{$>$}\llap
     {\lower.5ex\hbox{$\sim$}}}\ht1=\grdimen\dp1=0pt
\setbox\simlessbox=\hbox{\raise.5ex\hbox{$<$}\llap
     {\lower.5ex\hbox{$\sim$}}}\ht2=\grdimen\dp2=0pt

%
% from Larry Molnar
% Set up some definitions:
%
%This is how to have an approximate sign under < or > :

\def\ref#1  {\noindent \hangindent=24.0pt \hangafter=1 {#1} \par}
\def\doublespace {\smallskipamount=6pt plus2pt minus2pt
                  \medskipamount=12pt plus4pt minus4pt
                  \bigskipamount=24pt plus8pt minus8pt
                  \normalbaselineskip=24pt plus0pt minus0pt
                  \normallineskip=2pt
                  \normallineskiplimit=0pt
                  \jot=6pt
                  {\def\smallskip {\vskip\smallskipamount}}
                  {\def\medskip   {\vskip\medskipamount}}
                  {\def\bigskip   {\vskip\bigskipamount}}
                  {\setbox\strutbox=\hbox{\vrule
                    height17.0pt depth7.0pt width 0pt}}
                  \parskip 12.0pt
                  \normalbaselines}
\magnification=1200
$$\bf NON-THERMAL\ ACCELERATION\ FROM\ RECONNECTION\ SHOCKS$$
$$\rm Eric\ G.\ Blackman\ and\ George\ B.\ Field$$
$$\rm Harvard-Smithsonian\ Center\ for\ Astrophysics$$
$$\rm 60\ Garden\ St.,\ Cambridge\ MA,\ 02138$$
\doublespace
$$\bf ABSTRACT$$

Reconnection shocks in a magnetically dominated plasma must be
compressive.  Non-thermal ion acceleration
can occur across built-in slow shocks, and across outflow fast shocks
when the outflow is supermagnetosonic and the
field is line-tied.  Electron acceleration may be initiated by
injection from the dissipation region.  Reconnection and
shock acceleration thus cooperate and non-thermal acceleration
should be a characteristic feature.
\vfill
\eject

Magnetic reconnection converts magnetic energy into particle energy as
oppositely magnetized flows merge across a thin dissipation region (DR).
An upper limit to the (2-D)
reconnection rate comes from the Petschek
model (PK)
originally proposed to explain rapid solar flare energy release [1].
In PK, the DR length $<<$ field gradient length and
magnetic tension in the outflow thrusts the plasma from the DR.
Slow shocks are built explicitly into PK,
across which the plasma flow changes abrubtly
from inflow to outflow.  The PK tension
thrust provides faster reconnection than the Sweet-Parker (SP)
type models for which the DR length $\sim$ field gradient length [2].

Though slow shocks are built into PK, similar structures are seen numerically
even when the DR $>>$ than that of PK [3].  In addition, fast shocks can be
present when the
outflow is supermagnetosonic and the outflow boundary field is kept
fixed [4].
Although real astrophysical plasmas are compressive,
most work has employed the incompressible
approximation to facilitate a global solution.
Compressible reconnection models have
also focused on perturbative solutions and the rate [5], not
on the particle spectrum.
Here we consider the effect of compressive shocks near a DR on the spectrum.

Slow shocks, unlike fast shocks, have not been extensively simulated,
and their potential for power law acceleration has not been
addressed as an important outcome of reconnection.
The shocks may accelerate ions directly from a thermal
distribution and the DR may provide the injection
electrons required for non-thermal electron acceleration.
Since much of the flow of a reconnecting region passes through
the shocks, reconnection may possibly explain sustained non-thermal
features in magnetically dominated astrophysical phenomena.

We first solve the jump conditions across the built-in slow shock for the
compression ratio, given a magnetically dominated inflow.
We find the parameter space for which the
outflow must be supermagnetosonic and relate the
the compression ratio across the slow shock, $r_s$, to that across the
fast one, $r_f$.  We then discuss the implications for shock acceleration.

The non-relativistic MHD jump conditions for mass, momentum, and energy are [6]
$$\rho_1v_{1n}=\rho_2v_{2n},\eqno (1)$$
 $$\rho_1 v_{1n}^2+P_1+B_{1t}^2 / 8 \pi=\rho_2
v_{2n}^2+P_2+B_{2t}^2/8\pi,\eqno(2)$$
$$\rho_1v_{1n}{\bf v_{1t}}-B_{1n}{\bf B_{1t}}/4\pi
=\rho_2v_{2n}{\bf v_{2t}}-B_{2n}{\bf B_{2t}}/4\pi,\eqno(3)$$
$$(1/2)\rho_1v_1^2v_{1n}+\Gamma(\Gamma-1)^{-1}P_1v_{1n}+(B_1^2/4\pi)v_{1n}-{\bf
v_1\cdot B_1}B_{1n}/4\pi
=(1/2)\rho_2v_2^2v_{2n}$$
$$+\Gamma(\Gamma-1)^{-1}P_2v_{2n}+(B_2^2/4\pi)v_{2n}-{\bf v_2\cdot
B_2}B_{2n}/4\pi,\eqno(4)$$
where $B$ is the magnetic field, $v$ is the velocity,
$P$ is the pressure, $\rho$ is the density and $\Gamma$ is the adiabatic
index.  The subscript $1(2)$ refers to the up(down)stream
region, and the subscript $n(t)$ refers to the normal (tangential)
components.
The electromagnetic jump conditions for an ideal plasma are given by
$$ B_{1n}=B_{2n}\eqno(5)$$
$$({\bf v_1}{\rm x}{\bf B_1})=({\bf v_2}{\rm x}{\bf B_2}).\eqno(6)$$
The shock is $\perp$ to the ${\hat {\bf n}}, {\hat {\bf y}}$
plane as shown in Fig. 1.  We assume the switch-off condition,
$B_{2y}=0$, and also that ${\bf v}_1/|v_1|\cdot {\hat{\bf y}}<< 1$ [8].
Define ${\tilde c}\equiv cos\theta$, ${\tilde s} \equiv sin\theta$
and ${\tilde t}\equiv tan\theta$ where $\theta$ is the
angle between the downstream flow and the shock normal.
Define $c_{1} \equiv cos {\phi}_{1}$, $s_{1}
\equiv sin {\phi}_{1}$
and $t_{1} \equiv tan { \phi}_{1}$  where $  \phi_{1}$
is the angle between the upstream field and the shock normal.
The configuration of Fig. 1 is then described by
$$v_{1n}=-v_1 ,\ B_{1n}=-B_1 c_1,\ B_{1y}=-B_1 s_1,
v_{2n}=-v_2 {\tilde c},\ v_{2y}= v_2 {\tilde s},\ B_{2n}=-B_2,\
v_{1y}=B_{2y}=0. \eqno(7)$$
For $\Ga=5/3$ and $\beta_1 \equiv a_{1s}^2/v_{1A}^2 <<1$, where
$a_{1s}$ and $v_{1A}$ are the inflow sound
and Alfv\'en speed, plugging $(7)$ into $(1)$-$(6)$ gives
$$t_1^2=2(r_s-1)(r_s-4)/(5r_s-2r_s^2),\eqno(8)$$
$$\beta_2=(5/3)[(r_s-1)/r_s+t_1^2/2]\eqno(9)$$
$$M_{2A}^2\equiv v_2^2/v_{2A}^2=(1+r_s^2t_1^2)/r_s,\eqno(10)$$
where $M_{2A}$ is the outflow Mach number, $v_{2A}$ is the outflow
Alfv\'en speed, $\beta_2\equiv a_{2s}^2/v_{2A}^2$,
and $a_{2s}$ is the outflow
sound speed.  Fig. 2 shows $t_1$ and $M_{2A}$ versus $r_s\equiv \rho_2/\rho_1$.
Since $t_1^2>0$,
$(8)$ shows that $2.5<r_s<4$ for a low
$\beta_1$ switch-off shock [7], with the lower limit being a
perpendicular $(\perp)$ shock and the upper limit a parallel $(||)$ shock.
As $\beta_1\to\infty$, $r_s \to 1$.

When ${\bf B}_2\cdot {\bf v}_2/|B_2v_2|<< 1$, $v_2$
will be supermagnetosonic [6] when
$v_{2}^2=v_{2n}^2+v_{2y}^2>a_{2s}^2+v_{2A}^2$.
Using $(7)$, $(2)$, $(3)$, and $(5)$ this condition reduces to
$6r_s^2-13r_s-20 < 0,$
and is satisfied for
$r_s<3.2$ or $t_1>1.25$ from  $(8)$.
A nearly uniform supermagnetosonic outflow becomes the condition for
a fast shock when the field is line-tied at the outflow boundary.
The jump conditions, $(1)-(6)$, across such a
quasi-$\perp$ fast shock for $\Ga=5/3$ give the equation
$$M_{2A}^2=3r_f\beta_2/(4-r_f)+(3/2)r_f(r_f-1)/(4-r_f).\eqno(11)$$
Combining this with $(8)$, $(9)$, and $(10)$ we obtain,
$$r_f=(12r_s-30)^{-1}[4r_s^2-20r_s+7+(16r_s^4-544r_s^3+2952r_s^2-4312r_s-431+
2400/r_s)^{1/2}].\eqno(12)$$
Fig 2. shows that $1<r_f<2$ when $3.2>r_s>2.5$.
The inverse dependence is expected because a decrease in $r_s$
 corresponds to an increase in tension force along the
shock plane, and thus a larger $M_{2A}$, accounting for the
larger $r_f$.  For the canonical quasi-$\perp$
built-in slow shock with a line-tied
outflow, a fast shock is likely,
as the supermagnetosonic outflow condition requires only that $t_1>1.25$.

The compression ranges of $2.5<r_s<4$ for the built-in slow
shock and $1<r_f<2$ for the fast shock
are important for shock acceleration theory:  For
distribution functions isotropic to first order in
$v/v_p^*$ where $v_p^*$ is the particle velocity in the proper frame
of the bulk flow $v$, the steady state Boltzmann equation can be written
as a diffusion-convection (DC) equation.  We define
$N(x,p_p)dp_p\equiv 4\pi p_p^2 f(x,{p_p})dp_p$, where
$f$ is the Boltzmann distribution function, $x$ measures position, and
$p_p$ is the particle momentum.  The DC equation across a general
shock is then [9]
$$\partial_n[v_nN-\ka_n \partial_nN]-
(1/3)(\partial_n v_n)\partial_{p_p}[p_pN]=0,\eqno(13)$$
where $v_n$ is the normal flow velocity, $\ka_n$ is the normal diffusion
coefficient, and we have assumed
that gradients in the normal direction $>>$ those along the shock.
The solution of $(13)$ across the shock when the shock thickness $<<$
mean free path [9] shows that the outflow energy spectrum for a
steeper inflow  spectrum takes the
power law form $N\propto p^{-w}_p$ with energy index $w=(r+2)/(r-1)$
depending only on the compression ratio, $r$.
Fermi acceleration operates as the
particles diffuse between scattering centers (presumably turbulence)
on each side of the shock.  Particles always
see the centers converging, as the normal velocity is larger upstream.

Shock acceleration can dominate
synchrotron loss when $\tau_{syn}$, the shortest
synchrotron loss timescale of the region, exceeds
the longest shock acceleration timescale $\tau_{sh}$:
$$\tau_{syn}\equiv 6\pi m_ec/(\ga_e B_1^2 \si_T)>\tau_{sh}\sim
\kappa_{n||}/v_1^2
,\eqno(14)$$
where $\si_T$ is the Thomson cross section, $\ga_e$ is the electron
Lorentz factor, and $\kappa_{n||}$ is the
diffusion coefficient normal to the slow shock and thus $||$ to the
downstream field.  For particles moving at $c$, $\ka_{n||}\sim
c\la_{||}/3$, where $\la_{||}$ is the field gradient
length [9] which we assume is of the same order in the inflow and
outflow regions.  From $(14)$, the condition justifying the absence of a
synchrotron loss term in $(13)$ is then
$$\gamma_e<\sim .08(v_1/{\rm\
cm\ sec^{-1}})^2(B_1/{\rm Gauss})^{-2}(\lambda_{||}/{\rm cm})^{-1}. \eqno(15)$$

The third term in $(13)$ can be thought of as the 1st order correction
to the rest frame equation, when measured in the lab frame.  This
motivates the finding [10] that $(13)$ includes not only guiding center
diffusion through pitch angle scattering,
but also drift acceleration from motion along the induced electric field.
The per particle energy change from the latter increases with
obliquity.
For slow (fast) shocks, the curvature (gradient) drift is
$||$ to the electric field and accounts for
energy gains while the gradient (curvature)
drift is anti-$||$ to the electric
field and incurs particle energy losses [11].  These contributions
conspire with those from the gyromotion component along the electric
field for a net per particle momentum change $dp_p/dt=-p_p\nabla^{\bf
.}v_\perp$, where $v_\perp$ is the flow velocity $\perp$ to $\bf B$.
The relevant component of this force is included in the coefficient of the
$\partial N/\partial p_p$ term of $(13)$.

The relative importance of a reconnection fast shock varies
inversely with the size of the DR:  When both shocks are present
$3.2>r_s>2.5$ with $2.4<w_s<3$, so that $1<r_f<2$ with $w_f \ge 4$,
where $w_{s(f)}$ is the slow (fast) shock energy index.
(The range $3.2 \le r_s<4$ corresponds to a slow shock with no outflow
fast shock and $2.4\ge w_s>2$.)
Thus, if slow shock acceleration is effective,
fast shocks cannot further steepen the already steep spectrum
of particles that passed through
the slow shock.  However, if the DR length $\sim \la_{||}$, then
more of the flow will see only the fast shock, and the
spectrum from shock acceleration
should have a somewhat lower energy index.

Although we have used a DC equation,
we recognize that shock acceleration is a rather
non-linear process.  However, fast shock simulations show that the Fermi
acceleration engine is very efficient, transferring $\ge 1/10$ of the
inflow energy to particles [9].  These
particles tend to smooth out the
shock by diffusion, produce turbulence,
and $increase$ the compression ratio above the
jump condition value, as their escape and acceleration
change the downstream equation of state.
The shock smoothing can
violate the assumptions built into the simple DC scheme, but
the increase
in the compression ratio over the linear limit $enhances$ the
non-thermal acceleration.
The efficiency is relatively
insensitive to the obliquity of the inflow field
unless the Mach number exceeds $\sim 30$.
For our case, outflow fast shocks would then be effective if $M_{2A}\sim
r_st_1<30$ from $(10)$.

A recent hybrid simulation [12] of oblique slow shocks and the ion-ion
cyclotron instability shows that steady or cyclically reforming slow shocks
can develop depending on the inflow conditons.  In the steady case a
coherent Alfven wave train forms downstream.  In the unsteady case
the shock quasi-periodically transforms from a thin sharp transition
to a wide diffuse transition, and Alfv\'en turbulence is seen
downstream.  This cyclic reformation is strikingly similar to that
seen in quasi-$||$ fast shocks [9] and is an example of an
electromagnetic beam instability
brought on by the interaction of backstreaming ions with the inflowing
plasma.  Waves produced in the upstream by such instabilites are
amplified as they convect back to the shock front.  The compressed
waves then interact strongly with the inflow particles, scattering and slowing
them, producing the entropy required for the shock.  Some of the ions
are scattered back upstream by the waves, and a subset of those
are scattered back to the shock.  The reformation of the thin structure
occurs if the backstreaming particles leave the simulation region, and
then no longer produce waves that are convected to the shock.
The above process is in fact how
non-thermal ions are extracted from an initially thermal input,
initiating the Fermi process.  It is the effectiveness of the Fermi
acceleration which makes shock acceleration such a non-linear process.

Slow shocks in the geomagnetic tail show wave substructures with
properties similar to those of Ref. [12], and also show turbulence ahead and
behind the shock fronts [13].  In addition, although much of the
shock acceleration goes into ions, significant non-thermal
tails in the electron spectra are seen [14], and are not modeled by hybrid
simulations which assume a fluid electron population.  Fast shock simulations
show electron acceleration with injection electrons [15], and
we expect that slow shocks could operate similarly.  More simulations
of slow shocks are needed which predict the spectrum of accelerated
particles.

Jet plasma in AG may be largely pair plasma [16], so ion-electron simulations
might not be applicable.
We suggest that reconnection and its shocks are a strong
candidate for solving the re-acceleration problem
in jets, which requires sustaining non-thermal electron emission over distances
exceeding $c\tau_{syn}$, where $c$ is the speed of light.  Shock electron
acceleration requires injection particles [16] which we now
show, and argue that the DR may be able to provide such injection.

We can estimate the range of
particle energies accelerated by the Fermi process for the slow and
fast shocks, and in particular the range of $\ga_e$.
First consider the slow shock:
An upper limit can be found by
 ensuring that the particles see an ordered field.  This requires
$\lambda_{||}>g_2$ where $g_2\sim c^2\gamma_em_e/(eB_2)$ is the particle
gyroradius associated with the smaller field of the two flow regions.  This
implies $\gamma_e<eB_2\la_{||}/(m_ec^2)\sim
10^{-3}(B_2/{\rm Gauss})(\la_{||}/{\rm cm})$.
A lower limit can be found by demanding that
the downstream particles be able to diffuse upstream.  This requires
particles of large enough energy to resonantly
interact with the plasma waves which pitch-angle scatter the particles
upstream.  For Alfv\'en turbulence [16], the electron lower bound is a factor
$\sim m_p/m_e$ times that for protons and is given by
$\ga_e>(m_p/m_e)(v_{2A}/c)$ where $m_p$ is the proton mass
and $e$ is the charge.  Thus for the slow shock
$$(m_p/m_e)(v_{2A}/c)<\ga_e<10^{-3}(B_2/{\rm Gauss})(\la_{||}/{\rm
cm}).\eqno(16)$$

Consider now the fast shock:
Since the flow downstream from the fast shock is primarly $\perp$ to the shock
normal, diffusion across the shock requires [9]
$$\ka_{n\perp}/v_3>g_3\sim \gamma_em_ec^2/(eB_3),\eqno(17)$$
where $v_3$
is the downstream flow speed, $B_3$ is the downstream field, $g_3$ is
the associated gyroradius, and  $\ka_{n\perp}$  is the diffusion coefficient
normal to the shock but $\sim \perp$ $\bf B_3$.
For $\la_{||}>g_3$ we use $\ka_{n||}\sim c\la_{||}/3$, so [9]
$\ka_{n\perp}=cg_3^2/(3\la_{||})$ and
$(17)$ gives $\ga_e>3 e B_3 v_3 \la_{||}/(m_e c^3)$.
This limit must be combined with the analogous limits as described for
the slow shock so that for the fast shock we have
$${\rm Max}\ [(m_p/m_e)(v_{3A}/c),\
10^{-3}(B_3/{\rm Gauss})(\la_{||}/{\rm cm})(v_{3}/c)]<\gamma_e<
10^{-3}(B_2/{\rm Gauss})(\la_{||}/{\rm cm}),\eqno(18)$$
where $v_{3A}$ is the Alfv\'en speed downstream from the fast shock.
Note that the upper limits in $(16)$ and
$(18)$ are less than the upper limit in $(15)$ when
$v_1^2>(.01{\rm cm \cdot sec}^{-2} {\rm Gauss}^{-3})B_1^2B_2\la_{||}$.

The highly dissipative energy conversion in the DR
could provide the injection electrons.  To see this,
note that upon absorbing the annihilated
field energy, the average $\gamma_e$ there $\sim(v_{1A}^2/c^2)(m_p/2m_e)$.
Combining this
with $(17)$ we see that slow shocks of large obliquity favor a DR
which can inject, as the condition is
$$v_{1A}/v_ {2A}=r_s^{1/2}/c_1>\sim c/v_{1A}.\eqno(19)$$

We have discussed non-thermal acceleration by low $\beta_1$
reconnection slow shocks and
outflow fast shocks.  The particle spectra across the slow shocks
should be at least as flat as that given by the DC spectral index range
of $2<w_s<3$ for sufficiently large slow shocks.
The dissipation region can provide injection electrons
for acceleration and the above index range is consistent with observed
features of radio galaxy lobes and jets [16].

-----------------------------

\noindent [1] H. E. Petschek, in $Physics\ of\ Solar\ Flares$,
$AAS-NASA\ Symposium$, NASA SP-50, W. N. Hess, ed., p. 425 (1964).

\noindent [2] P. A. Sweet, Proc. IAU Symp. No. $\bf 6$, 123 (1958);  E. N.
Parker, J. Geophys. Res., $\bf 79$, 1558 (1957).

\noindent [3] D. Biskamp, Phys.  Rep., $\bf 237$, 179 (1994).

\noindent [4] T. G. Forbes, Ap. J., $\bf 305$, 553 (1986).

\noindent [5] M. Jardine \& E. R. Priest, in $Reconnection\ in\ Space\
Plasma$, edited by T. D. Guyenne \& J. J. Hunt, (ESA SP-285, ESCTEC,
Noorwidjk, Netherlands, 1989); T. Sato, J. Geophys. Res., $\bf 84$,
7177 (1979).

\noindent [6] D. B. Melrose, $Instabilities\ in\ Space\ and\
Laboratory\ Plasmas$ (Cambridge Univ. Press, Cambridge, United
Kingdom, 1986).

\noindent [7] A. Kantrowitz \& H. E. Petschek, in $Plasma\ Physics\
in \ Theory\ and\ Application$
edited by W. B. Kunkel, (McGraw-Hill, New York, 1966).

\noindent [8] If this were not the case, we could translate
along the shock to make it so, since the compression
ratio is invariant.  We would then translate the resulting outflow
velocity back to the lab frame to test the supermagnetosonic condition.

\noindent [9] F. C. Jones \& D. C. Ellison, Space Sci. Rev., $\bf 58$, 259
(1991).

\noindent [10] F. C. Jones, Ap. J., $\bf 361$, 162 (1990).

\noindent [11] G. M. Webb, W. I. Axford \& T. Terasawa, Ap. J., $\bf
270$, 537 (1983).

\noindent [12] N. Omidi \& D. Winske, J. Geophys. Res., $\bf 97$, 14801 (1994).

\noindent [13] F. V. Coroniti et al., J. Geophys. Res., $\bf 99$, 11251 (1994).

\noindent [14] W. C. Feldman et al., J. Geophys. Res., $bf 90$, 233 (1990).

\noindent [15] D. C. Ellison, in $Particle\ Acceleration\ in\ Cosmic\
Plasmas$, edited by G. P. Zank \& T. K. Gaisser (American Institute of
Physics, New York, 1992).

\noindent [16] J. A. Eilek \& P. A. Hughes, in $Beams\ and\
Jets\ in\ Astrophysics,$ edited by P. A. Hughes,
(Cambridge Univ. Press, Cambridge 1991).

\end